\documentclass[aps,prl,reprint,superscriptaddress]{revtex4-2}
\usepackage{graphics}
\usepackage{graphicx}

\begin{document}

% Use the \preprint command to place your local institutional report
% number in the upper righthand corner of the title page in preprint mode.
% Multiple \preprint commands are allowed.
% Use the 'preprintnumbers' class option to override journal defaults
% to display numbers if necessary
%\preprint{}

%Title of paper
\title{Failure of LMC statistical complexity in identifying structural order in the XY model}

% repeat the \author .. \affiliation  etc. as needed
% \email, \thanks, \homepage, \altaffiliation all apply to the current
% author. Explanatory text should go in the []'s, actual e-mail
% address or url should go in the {}'s for \email and \homepage.
% Please use the appropriate macro foreach each type of information

\author{Dario Javier Zamora}
\email[]{ djzamora@conicet.gov.ar}
\affiliation{Instituto de Fisica del Noroeste Argentino, CONICET and UNT. Av. Independencia 1800, Tucuman, CP 4000, Argentina.}

\date{\today}

\begin{abstract}
Quantifying complexity in physical systems remains a fundamental challenge, and many proposed measures fail to capture the structural features that intuitive or theoretical considerations would demand. Among them, the López-Ruiz–Mancini–Calbet (LMC) statistical complexity has been widely cited due to its simplicity and analytic tractability. Here, we examine the performance and limitations of the LMC measure in a controlled physical setting: a two-dimensional XY model studied through Monte Carlo simulations. By computing LMC complexity at each step of the system’s relaxation dynamics, and directly comparing these values with the evolving dipole configurations, we show that LMC complexity systematically fails to identify states of high structural complexity. In particular, the measure often assigns maximal complexity to nearly equilibrated configurations while underestimating vortex-rich intermediate states. We further show that the time derivative of LMC complexity contains more meaningful dynamical information. We propose that future measures incorporate directionality and dynamical sensitivity to better reflect the emergence and decay of organization in nonequilibrium systems.
\end{abstract}

% insert suggested keywords - APS authors don't need to do this
%\keywords{}

%\maketitle must follow title, authors, abstract, and keywords
\maketitle

% body of paper here - Use proper section commands
% References should be done using the \cite, \ref, and \label commands
\section{\label{sec:level1}Introduction}

A complex system is an entity composed of multiple interrelated or interdependent components that, as a whole, exhibit properties and behaviors that are not evident in the individual parts \cite{Scott2005}. The components of a complex system are often connected through non-trivial networks, which may include both local and long-range interactions. The structure of these connections can significantly influence the system’s overall behavior. Complex systems may exhibit phase transitions and critical points, where small changes in parameters lead to drastic changes in the system’s behavior \cite{Bar-Yam1997}.

The study of complex systems focuses on understanding how collective properties emerge from the interactions among components and how these systems can be modeled, analyzed, and, in some cases, controlled \cite{Boccara2010}. Two key features of complex systems are:
(i) Emergence, which refers to the generation of new and often unexpected macroscopic properties and patterns arising from microscopic-level interactions—behaviors not explicitly determined by the rules governing the individual components; and
(ii) Self-organization, the capacity of a system to spontaneously develop ordered structures and patterns from disordered initial conditions, without external intervention or centralized control \cite{Thurner2018}.

Contemporary thought suggests that natural irreversible processes often exhibit self-organization, leading to the spontaneous emergence of complexity. The study of complexity can be approached through multiple avenues, and numerous proposals have been developed. In standard mathematical models of many-body systems—such as stochastic cellular automata, the Ising model, and partial differential equations used in hydrodynamics or reaction–diffusion systems—researchers have sought to identify general principles that govern the creation and destruction of complexity. A central challenge in this effort is to develop definitions of complexity that are both mathematically rigorous and consistent with intuitive understanding. Such definitions must be precise enough to support formal theorems, yet sufficiently broad to meaningfully capture the diverse phenomena classified as complex. The question of how to define complexity has received considerable attention in the literature and remains an active topic in contemporary science.

Many definitions of complexity have been proposed, including algorithmic complexity \cite{Kolmogorov1968}, logical depth \cite{Bennett1995}, effective complexity \cite{Gell-Mann1996}, thermodynamic depth \cite{Lloyd1988}, and statistical complexity \cite{Feldman1998}. However, there is currently no universally accepted definition. A major limitation of these proposals is that each captures only a specific facet of complexity. Moreover, some are not computable, while others lack formal rigor. This proliferation of measures has introduced a degree of confusion within the field of complex systems. A non-exhaustive list of such measures can be found in \cite{Bennett2002} and \cite{Lloyd2001}.

A general intuitive notion of complexity refers to the degree to which a system’s components are coupled into organized structures. High complexity is often associated with a balance between randomness and regularity, enabling the emergence of new phenomena and self-organized behavior. While there is broad agreement on this qualitative interpretation, a fundamental open question in complex systems theory is how to formulate a quantitative definition of complexity.

It is generally expected that a measure of complexity reaches its maximum at intermediate levels of entropy and tends toward zero for both highly ordered and highly disordered systems. Neither a fully regular nor a completely random pattern is typically perceived as complex. Rather, complexity tends to arise in structures that lie between these two extremes, where order and disorder coexist in a dynamic balance.

One of the most widely discussed statistical measures of complexity is the LMC measure, introduced by López-Ruiz, Mancini, and Calbet \cite{Lopez-Ruiz1995}. The LMC measure is defined as 

\begin{equation}
C = H \cdot D,
\label{def}
\end{equation}

where $C$ is the complexity, $H$ is the Shannon entropy, and $D$ is the disequilibrium, a measure of how far a probability distribution deviates from uniformity.

The Shannon entropy is defined as

\begin{equation}
H = -K \sum_{i=1}^N p_i \log p_i,
\end{equation}

and the disequilibrium as

\begin{equation}
D = \sum_{i=1}^N \left(p_i - \frac{1}{N}\right)^2,
\label{diseq}
\end{equation}

where $p_i$ denotes the probability associated with state (i), and the distribution spans $N$ possible states. The constant $K$ depends on the unit system, with common choices being $K = 1$ or $K = k_B$, the Boltzmann constant. It is intended to capture the interplay between the amount of information contained in a system and the distance of its probability distribution from equipartition (uniformity).

This construction ensures that the complexity $C$ vanishes in both limiting cases of perfect order and complete disorder—an essential (though not sufficient) property of a meaningful complexity measure. In highly ordered states, such as crystal-like configurations, the entropy $H$ approaches zero. In contrast, in maximally disordered (e.g., gas-like) systems, the disequilibrium $D$ approaches zero. Hence, the product $C = H \cdot D$ is zero in both extremes.

The previous equations can be generalized to continuous probability distributions by replacing the summations with appropriate integrals. Additional generalizations have also been proposed, involving alternative divergence measures (Kullback-Leibler divergence for example \cite{Feldman1998}), as well as extensions based on non-standard entropies, such as Tsallis entropy \cite{Kowalski2012}.

In this paper, I present a study of the behavior of the LMC measure, arguing that—although the LMC measure exhibits some desirable features—it does not fully align with the intuitive notion of complexity. To do so, I use an XY model studied through Monte Carlo simulations. I compute entropy and other complexity-related measures step by step, allowing for a detailed and visual comparison with qualitative notions of complexity. The XY model exhibits the necessary characteristics for such an analysis and is historically significant in physics. Furthermore, it is flexible and extensible, allowing for the introduction of more complex features such as glassy behavior, changes in dimensionality, frustration, long-range interactions, and varying temperature.

\section{Methods}
The XY model represents a system of magnetic dipoles that can orient freely within a plane, with each dipole characterized by an angle ranging from $0$ to $2\pi$. In this simulation, the spins are arranged in a two-dimensional $15 \times 15$ lattice. This lattice size offers a practical balance: it is small enough to ensure reasonable computational performance, yet sufficiently large to allow observable ordering effects to emerge. Periodic boundary conditions are applied to mitigate edge effects.

The simulation code allows for the initialization of system parameters such as temperature, lattice size, and the initial spin configuration, which is randomized in all simulations. The Metropolis algorithm is implemented using the sweep method, where each spin is sequentially perturbed. Each proposed orientation change is accepted or rejected based on the change in system energy and the temperature, according to the Boltzmann probability distribution. This reflects the physical principle of energy minimization in thermal equilibrium.

The interaction energy $E_{ij}$ between two nearest-neighbor spins $s_i$ and $s_j$ is defined as:

\begin{equation}
E_{ij} = -\mathbf{s}_i \cdot \mathbf{s}_j = -\cos(\theta_i - \theta_j).
\end{equation}

Thus, the energy is minimized ($E_{ij} = -1$) when the spins are aligned, i.e., when $\theta_i = \theta_j$, representing ferromagnetic order.

Our code computes the total energy of the system, as well as thermodynamic quantities such as entropy and the angular probability distribution function (PDF). These values are evaluated over time as the system evolves toward equilibrium. At each time step, the Metropolis algorithm yields a histogram of the angle distribution, denoted by $p(\theta)$. From this distribution, the Shannon entropy is calculated as:

\begin{equation}
H = -\int_{0}^{2\pi} p(\theta)\cdot\ln\left(p(\theta)\right)\cdot d\theta.
\end{equation}

This entropy measure provides insight into the system’s degree of disorder and is a key component in evaluating its statistical complexity.

In Figs.~(\ref{key}) and (\ref{fig:LMCsimulation}), we present an example of a simulation at $T = 0.01 K$. Initially, the spins are randomly oriented, meaning that each angle $\theta$ is equally likely. In other words, the probability distribution is uniform: $p(\theta) = p = \text{const}$.

At the other extreme, if the system evolves long enough at low temperatures, all dipoles eventually align. In this case, the angular distribution approaches a Dirac delta: one specific orientation has probability 1, and all others have probability 0. As a result, entropy reaches its minimum, and a fully ordered state is achieved.

The most interesting behavior occurs at intermediate stages of the simulation, when structured patterns begin to emerge. At these stages, entropy is neither maximal nor minimal—it takes intermediate values—and complexity reaches its peak. This is the regime in which topological structures such as vortices form (for example, as the one shown in the 4300-iteration snapshot). These configurations exhibit a mixture of order and randomness, often referred to as partial order. Such states are not easy to describe or recreate, as many dipoles must be oriented in highly specific directions. Moreover, this partial order is dynamic, meaning that it evolves over time.

\section{Results}

The qualitative notion of complexity suggests that completely ordered and completely disordered configurations both have minimal complexity, while configurations lying between these extremes are more complex.

At low temperatures, the system’s equilibrium state is ordered. This implies that, over time, entropy decreases while disequilibrium increases. %The evolution of entropy and disequilibrium for a typical simulation in which vortices appear are shown in \ref{Zamora2025}.

\begin{figure}[h!]
\centering
\includegraphics[width=\linewidth]{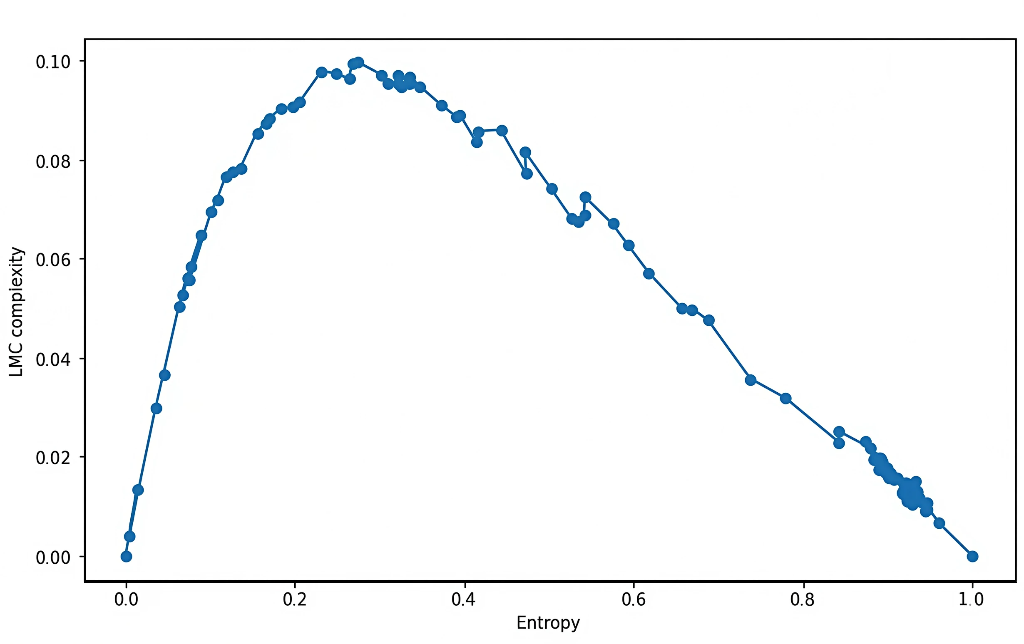}
\caption{LMC complexity vs entropy.}
\label{fig:LMCsimulation}
\end{figure}

The LMC measure captures the intuitive notion of complexity in this sense: it reaches a minimum for both fully ordered and fully disordered configurations, and a maximum for intermediate states, as illustrated in Fig.~(\ref{fig:LMCsimulation}) for this simulation.

\begin{figure*}
    \centering
    \includegraphics[width=\linewidth]{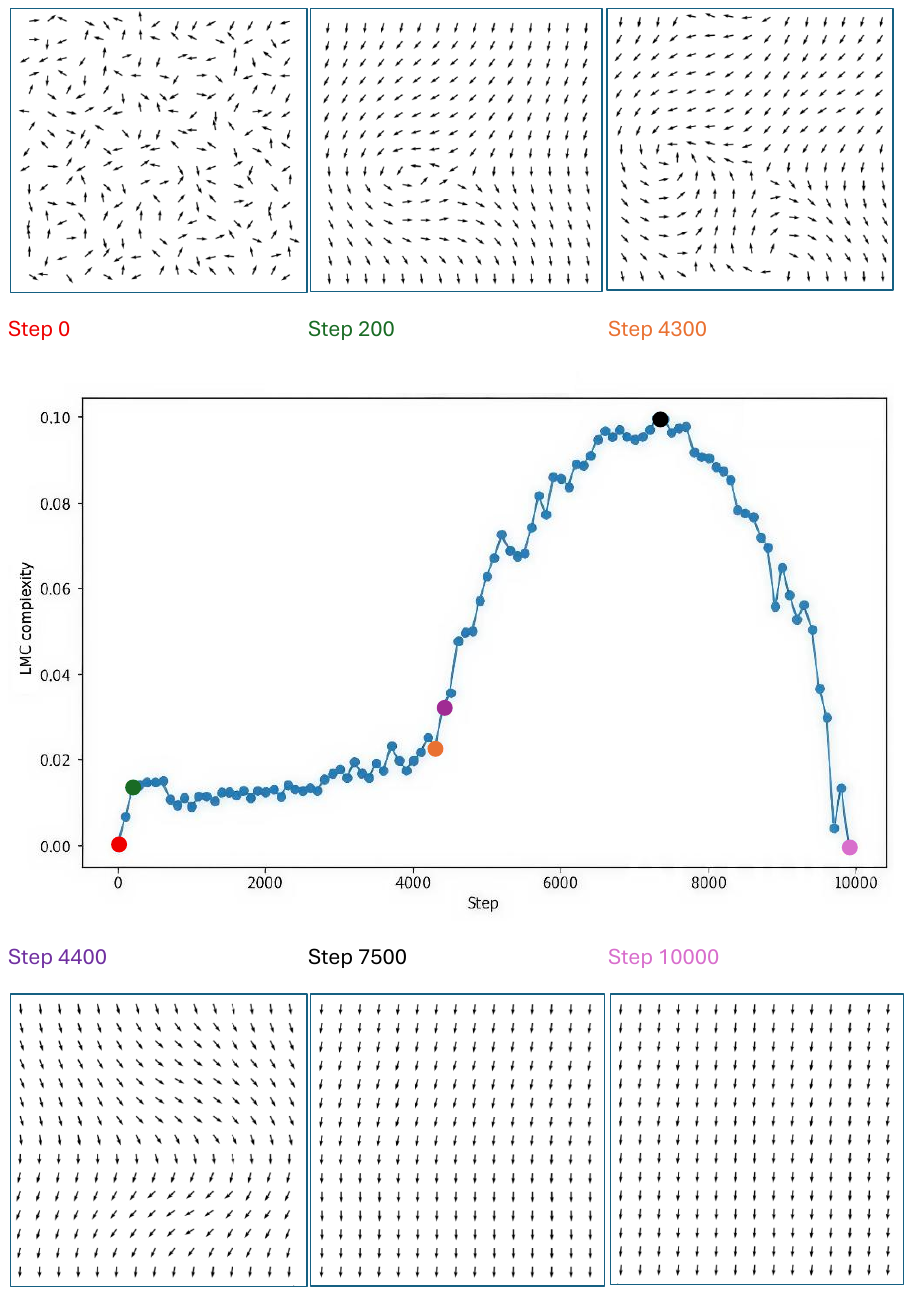}
    \caption{LMC complexity as a function of simulation step, together with the corresponding dipole configurations at six different times: steps 0, 200, 4300, 4400, 7500, and 10000. Each configuration is shown with a matching color label, and the corresponding point in the LMC-complexity curve is marked with the same color.}
    \label{key}
\end{figure*}

However, let us take a closer look to the configurations of spins at different stages of the simulation and compare them to the LMC complexity evolution in time. In Fig.~(\ref{key}), I present the dipole configurations at different stages of the simulation at $T = 0.01 K$, along with their corresponding LMC complexity values.

The curve shown corresponds to a specific simulation and does not represent a general behavior across all simulations. In other words, each time the simulation is run, the time evolution—and thus the functional form of the curve—varies. I selected this particular curve because it presents several distinct and interesting scenarios for analyzing how the LMC statistical complexity behaves.

First, note that the complexity is indeed zero for both the perfectly disordered and the perfectly ordered configurations (steps 0 and 10000, represented by the red and pink points, respectively). In this example, complexity increases rapidly during two distinct periods: at the beginning of the simulation and again around step 4300.

In the first phase (between steps 0 and 200), the increase in complexity corresponds to the spontaneous emergence of structure. The dipoles begin to rearrange from a random configuration, forming vortices and other nontrivial patterns. This behavior aligns well with the intuitive notion of complexity.

Following this phase, the LMC measure remains relatively stable until around step 4300, where a second sharp increase occurs. During this period, the vortices act like “knots,” leading to a metastable configuration that persists until thermal fluctuations eventually break the metastability. However, upon inspection of the dipole configurations at steps 4300 and 4400, it becomes clear that this second rise in complexity coincides with the disappearance of structure, not its formation. This contradicts the intuitive interpretation of complexity.

In fact, the configuration with the highest LMC complexity occurs at step 7500. Yet, when examining the dipole configuration at this step, we observe almost no structure—the system is nearly at equilibrium. This again suggests a disconnect between the LMC measure and our intuitive expectations.

In summary, the LMC measure does not reliably peak at the point where structural complexity is visually or physically most evident. This issue is not new; it was previously discussed by Anteneodo and Plastino \cite{Anteneodo1996}.

\section{Discussion}

LMC statistical complexity has the necessary (but not sufficient) property for a well-defined complexity measure of being zero in the states of maximum and minimum entropy. It also increases near points of instability, another desirable feature, since several observations indicate that complexity in nature often emerges close to instability points—such as the phase-transition points characteristic of critical phenomena \cite{Sole1997}.

LMC has been applied in several contexts, for example, in distinguishing between chaos and noise \cite{Olivares2012} (with limited success), detecting intermittency, and discriminating between distinct chaotic regimes \cite{Ferri2009}, as well as in studies of quantum–classical transitions \cite{Kowalski2005,Kowalski2007,Kowalski2011}.

The first problem with the LMC measure is related to its functional definition. The expression $C = H \cdot D$ does not work properly as a measure of complexity unless both $H$ and $D$ are normalized. Thus, the definition should be written as $C = \tilde{H} \tilde{D}$, where $\tilde{H}$ and $\tilde{D}$ are the normalized entropy and disequilibrium, respectively. In fact, this is how the concept of LMC complexity is applied throughout the literature for particular systems. What I want to emphasize here is that, due to this normalization procedure, the LMC measure is not universal in the sense that it cannot be used to compare different systems, but only different states of the same system, since entropy and disequilibrium are normalized to the maximum possible values within that specific system. This lack of universality poses a fundamental problem for any measure intended to quantify complexity, because complexity is intrinsically a cross-disciplinary concept applied to diverse domains of science—from physics and biology to economics and information theory. A meaningful complexity measure should, at least in principle, allow comparisons across different systems, models, or levels of description, and the LMC measure in its normalized form cannot satisfy this requirement.

The fact that LMC complexity increases when there are changes in the organization of the system indicates that the quantity $dC/dt$, let us call it "complexity production" for now, carries meaningful information about the evolution of the system’s structure, since it becomes larger when structures are created or annihilated. However, it would be desirable to have a measure whose time derivative becomes negative when structure is destroyed, thereby distinguishing between the creation and the loss of organization. 

I want to draw the reader’s attention to the fact that the quantity $dC/dt$ is, in a meaningful way, related to the concept of self-organization. This connection became clear when I read Bennett’s remark at the beginning of his paper \cite{Bennett2002}: “Natural irreversible processes are nowadays thought to have a propensity for \textbf{self-organization — the spontaneous generation of complexity}.” This observation suggests that changes in complexity—rather than its instantaneous value—may carry essential information about how structure emerges or dissipates within a system. In this sense, $dC/dt$ can be viewed as a dynamical indicator of the ongoing reorganization of the system's degrees of freedom.
From the definition of LMC complexity, equation \ref{def}, the complexity production is:

\begin{equation}
S = \frac{dC}{dt} = \frac{dH}{dt} \cdot D + \frac{dD}{dt} \cdot H,
\label{selforganization}
\end{equation}

where $H$ is the entropy, $D$ the disequilibrium, $\frac{dH}{dt}$ the entropy production, and, $S$ stands for "self-organization". Since $D$ represents a distance from the uniform probability distribution, $\frac{dD}{dt}$ can be interpreted as a kind of velocity toward (or away from) the uniform distribution. Notice that entropy production—an important quantity in irreversible processes and out-of-equilibrium physics—naturally appears in this expression.

From this perspective, the time derivative of complexity plays a role analogous to other rate-based quantities in physics, such as entropy production in nonequilibrium thermodynamics. These quantities do not merely characterize the state of the system but describe how that state evolves. Similarly, $dC/dt$ encapsulates how rapidly the system is moving toward or away from organized configurations, offering insight into processes such as structure formation, metastability, and relaxation.

This interpretation naturally motivates a broader question: to what extent can the balance between the two terms in Eq. (\ref{selforganization}) be understood as a type of complexity balance equation? The first term, $\frac{dH}{dt}\cdot D$, links entropy production to the system's deviation from uniformity, while the second term, $\frac{dD}{dt}\cdot H$, relates changes in structural heterogeneity to the current level of disorder. 

Regardless of the terminology—whether we call $dC/dt$ “complexity production,” “self-organization,” or something else—there are indications that this quantity contains valuable information about the dynamical state of the system. For this reason, I believe that $dC/dt$ deserves to be explored in much greater depth, both theoretically and in practical applications, as it may reveal aspects of complexity evolution that are not captured by static measures such as $C = H \cdot D$. For example, an important advance would be to modify Eq.~(\ref{selforganization}) in such a way that the sign of $dC/dt$ changes depending on whether structure is being created or destroyed, thereby introducing a notion of directionality into the measure.

After that, the complexity $C$, at the moment $\tau$, of a dynamical state can be defined as the cumulative of this quantity:

\begin{equation}
    C(\tau) = \int_0^\tau S(t) dt
\end{equation}

\section{Conclusions}
In this work, I examined the behavior and interpretive limitations of the LMC complexity measure when applied to a prototypical physical system, the XY model. Although the LMC measure possesses certain desirable properties—such as vanishing at states of maximum and minimum entropy and peaking at intermediate stages—it does not consistently align with intuitive or structural notions of complexity. Through explicit Monte Carlo simulations, I showed that LMC complexity may attain its highest values in configurations that are visually and physically simple, while assigning lower complexity to configurations rich in emergent structures such as vortices. This disconnect suggests that the LMC measure, in its original formulation, is not reliably capturing what physicists and complexity scientists typically understand as “complex.”

Nevertheless, the analysis presented here also points to a promising direction. While the instantaneous value of LMC complexity fails to reliably identify structurally complex states, its time derivative, $dC/dt$, does carry valuable dynamical information about the system. Changes in LMC complexity correlate with the creation or destruction of organization, suggesting that complexity may be better understood as a process rather than a static state.

Taken together, these findings suggest that the search for a single universal, static measure of complexity may be misguided—or at least insufficient. Complexity is a multifaceted concept, and different measures illuminate different aspects of how systems organize, store information, process interactions, and evolve over time. The LMC measure, despite its limitations, contributes to this broader toolkit, especially when its dynamical behavior is considered.

The question of whether complexity requires a rigorous mathematical definition remains open and deeply debated. On one hand, the ambition to quantify complexity has driven substantial progress in statistical physics, information theory, and nonlinear dynamics. Measures such as entropy, disequilibrium, logical depth, or statistical complexity attempt to capture specific aspects of what it means for a system to be complex. However, several authors have warned that the search for a single universal metric risks misunderstanding the nature of complexity itself.

As noted in \cite{Krakauer2024}, an overly narrow focus on numerical indicators turns complexity into a set of techniques—tools for identifying out-of-equilibrium patterns or extracting statistical structure from large datasets—rather than a framework for understanding the emergence of organization in natural and artificial systems. In the worst case, this tendency collapses the richness of the field into a misguided quest for a single scalar quantity capable of ordering systems as diverse as brains, societies, ecosystems, and technologies along one axis. As the cited work argues, such attempts resemble “trying to measure the degree of ‘chemistry’ in all chemical reactions or the quotient of ‘anthropology’ in a society.”

From this perspective, mathematical definitions—such as the LMC measure—remain valuable, but only insofar as they capture specific dimensions of complexity relevant to a given system. A single number can summarize certain behaviors, but cannot exhaust the richness of complex phenomena. Nevertheless, the attempt to define and measure complexity—whether through entropy-based measures like LMC, through computational metrics, or through dynamical indicators—remains valuable. The very effort to do so will surely enrich our understanding.
%3331 palabras con figuras y ecuaciones. Tiene que ser 3750 y 4 pag max+2 paginas de apendices

% Use the figure* environment if the figure should span across the
% entire page. There is no need to do explicit centering.

% Specify following sections are appendices. Use \appendix* if there
% only one appendix.
%\appendix
%\section{}

\begin{acknowledgments}
This research was financially supported by CONICET (National Scientific and Technical Research Council of Argentina). 
\end{acknowledgments}

\bibliography{xy}

\end{document}